\documentclass[12pt]{article}
\begin{document}
\title{A GAUGE LIKE FORMULATION OF GRAVITATION AND RELATED ISSUES}
\author{B.G. Sidharth\\
International Institute for Applicable Mathematics \& Information Sciences\\
Hyderabad (India) \& Udine (Italy)\\
B.M. Birla Science Centre, Adarsh Nagar, Hyderabad - 500 063 (India)}
\date{}
\maketitle
\begin{abstract}
After many fruitless decades of trying to unify electromagnetism and gravitation, it is now being realized that this can be done only in discrete spacetime, as indeed the author had demonstrated. In this context, a unified description of gravitation and electromagnetism is provided within the framework of a gauge like formulation. Following the discrete spacetime structure, we then argue that the underpinning for the universe is an array of Planck scale oscillators.
\end{abstract}
\section{Introduction}
In an earlier communication [1], based on a discrete spacetime noncommutative geometrical approach, we had shown that it was possible to reconcile electromagnetism and gravitation. It is of course well known that nearly ninety years of effort has gone in to get a unified description of electromagnetism and gravitation starting with Hermann Weyl's original Gauge Theory. It is only in the recent years that approaches in Quantum Gravity and Quantum Super Strings, amongst a few other theories are pointing the way to a reconciliation of these two forces. These latest theories discard the differentiable spacetime of earlier approaches and rely on a lattice like approach to spacetime, wherein there is a minimum fundamental interval which replaces the point space time of earlier theories. Indeed as Hooft has remarked, ``It is some what puzzling to the present author why the lattice structure of space and time had escaped attention from other investigators up till now....'' [2,3,4] Infact we had recently shown that within this approach, it is possible to get a rationale for the de Broglie wavelength and Bohr-Sommerfeld quantization relations as well[5]. Nevertheless, the link with the gauge theories of other interactions, based as they are, on spin 1 particles, is not clear, because the graviton is a spin 2 particle (or alternatively, the gravitational metric is a tensor).
\section{A Gauge like Formulation}
In this latter context, we will now argue that it is possible for both electromagnetism and gravitation to emerge from a gauge like formulation. In Gauge Theory, which is a Quantum Mechanical generalization of Weyl's original geometry, we generalize, as is well known, the original phase transformations, which are global with the phase $\lambda$ being a constant, to local phase transformations with $\lambda$ being a function of the coordinates [6]. As is well known this leads to a covariant gauge derivative. For example, the transformation arising from $(x^\mu) \to (x^\mu + dx^\mu)$,
\begin{equation}
\psi \to \psi e^{-\imath e \lambda}\label{e1}
\end{equation}
leads to the familiar electromagnetic potential
\begin{equation}
A_\mu \to A_\mu - \partial_\mu \lambda\label{e2}
\end{equation}
The above transformation, ofcourse, is a symmetry transformation. In the transition from (\ref{e1}) to (\ref{e2}), we expand the exponential, retaining terms only to the first order in coordinate differentials.\\
Let us now consider the case where there is a minimum cut off in the space time intervals. As is well known this leads to a noncommutative geometry (Cf.ref.[1])
\begin{equation}
[dx_\mu , dx_\nu ] = O(l^2)\label{e3}
\end{equation}
where $l$ is the minimum scale. From (\ref{e3}) it can be seen that if $O(l^2)$ is neglected, we are back with the familiar commutative spacetime. The new effects of fuzzy spacetime arise when the right side of (\ref{e3}) is not neglected. Based on this the author had argued that it is possible to reconcile electromagnetism and gravitation [7,8,9,10]. If in the transition from (\ref{e1}) to (\ref{e2}) we retain, in view of (\ref{e3}), squares of differentials, in the expansion of the function $\lambda$ we will get terms like
\begin{equation}
\left\{ \partial_\mu \lambda \right\} dx^\mu + \left(\partial_\mu \partial_\nu + \partial_\nu \partial_\mu \right) \lambda \cdot dx^\mu dx^\nu\label{e4}
\end{equation}
where we should remember that in view of (\ref{e3}), the derivatives (or the product of coordinate differentails) do not commute. As in the usual theory the coefficient of $dx^\mu$ in the first term of (\ref{e4}) represents now, not the gauge term but the electromagnetic potential itself: Infact, in this noncommutative geometry, it can be shown that this electromagnetic potential reduces to the potential in Weyl's original gauge theory [6,7].\\
Without the noncommutativity, the potential $\partial_\mu \lambda$ would lead to a vanishing electromagnetic field. However Dirac pointed out in his famous monopole paper in 1930 that a non integrable phase $\lambda (x,y,z)$ leads as above directly to the electromagnetic potential, and moreover this was an alternative formulation of the original Weyl theory [11,12].\\
Returning to (\ref{e4}) we identify the next coefficient with the metric tensor giving the gravitational field:
\begin{equation}
ds^2 = g_{\mu \nu} dx^\mu dx^\nu = \left(\partial_\mu \partial_\nu + \partial_\nu \partial_\mu \right) \lambda dx^\mu dx^\nu\label{e5}
\end{equation}
Infact one can easily verify that $ds^2$ of (\ref{e5}) is an invariant. We now specialize to the case of the linear theory in which squares and higher powers of  $h^{\alpha \beta}$ can be neglected. In this case it can easily be shown that
\begin{equation}
2 \Gamma^\beta_{\mu \nu} = h_{\beta  \mu ,\nu} + h_{\nu \beta ,\mu} - h_{\mu \nu ,\beta}\label{e6}
\end{equation}
where in (\ref{e6}), the $\Gamma$s denote Christofell symbols. From (\ref{e6}) by a contraction we have
\begin{equation}
2\Gamma^\mu_{\mu \nu} = h_{\mu \nu ,\mu} = h_{\mu \mu , \nu}\label{e7}
\end{equation}
If we use the well known gauge condition [13]
$$\partial_\mu \left(h^{\mu \nu} - \frac{1}{2} \eta^{\mu \nu} h_{\mu \nu}\right) = 0, \, \mbox{where}\, h = h^\mu_\mu$$
then we get
\begin{equation}
\partial_\mu h_{\mu \nu} = \partial_\nu h^\mu_\mu = \partial_\nu h\label{e8}
\end{equation}
(\ref{e8}) shows that we can take the $\lambda$ in (\ref{e4}) as $\lambda = h$, both for the electromagnetic potential $A_\mu$ and the metric tensor $h_{\mu \nu}$. (\ref{e7}) further shows that the $A_\mu$ so defined becomes identical to Weyl's gauge invariant potential [14].\\
However it is worth reiterating that in the present formulation, we have a noncommutative geometry, that is the derivatives do not commute and moreover we are working to the order where $l^2$ cannot be neglected. Given this condition both the electromagnetic potential and the gravitational potential are seen to follow from the gauge like theory. By retaining coordinate differential squares, we are even able to accommodate apart from the usual spin 1 gauge particles, also the spin 2 graviton which otherwise cannot be accommodated in the usual gauge theory. If however $O(l^2) = 0$, then we are back with commutative spacetime, that is a usual point spacetime and the usual gauge theory describing spin 1 particles.\\
We had reached this conclusion in ref.[1], though from a different, nongauge point of view. The advantage of the present formulation is that it provides a transparent link with conventional theory on the one hand, and shows how the other interactions described by non Abelian gauge theories smoothly fit into the picture.\\
Finally it may be pointed out that the author had argued that a fuzzy spacetime input explains why the purely classical Kerr-Newman metric gives the purely Quantum Mechanical anomalous gyromagnetic ratio of the electron [15,16], thus providing a link between General Relativity and electromagnetism. This provides further support to the above considerations.
\section{Planck Scale Considerations} 
Let us now push the Planck scale considerations further. In an earlier communication [17], we had argued that a typical
elementary particle like a pion could be considered to be the result of
$10^{40}$ evanescent Planck scale particles. The argument was based on 
random motions and also on the modification to the Uncertainity Principle.
We will now consider the problem from a totally different point of view,
which not only reconfirms the above result, but also enables an elegant
extension to the case of the entire universe itself.
Let us now consider an array of $N$ particles, spaced a distance $\Delta x$
apart, which behave like oscillators, that is as if they were connected by
springs. We then have [18,19]
\begin{equation}
r  = \sqrt{N \Delta x^2}\label{e1a}
\end{equation}
\begin{equation}
ka^2 \equiv k \Delta x^2 = \frac{1}{2}  k_B T\label{e2a}
\end{equation}
where $k_B$ is the Boltzmann constant, $T$ the temperature, $r$ the extent  and $k$ is the 
spring constant given by
\begin{equation}
\omega_0^2 = \frac{k}{m}\label{e3a}
\end{equation}
\begin{equation}
\omega = \left(\frac{k}{m}a^2\right)^{\frac{1}{2}} \frac{1}{r} = \omega_0
\frac{a}{r}\label{e4a}
\end{equation}
We now identify the particles with Planck masses, set $\Delta x \equiv a = 
l_P$, the Planck length. It may be immediately observed that use of 
(\ref{e3a}) and (\ref{e2a}) gives $k_B T \sim m_P c^2$, which ofcourse agrees 
with the temperature of a black hole of Planck mass. (Equations (\ref{e2a}) and (\ref{e3a}) define the spring constant $k$). Indeed 
Rosen had shown that a Planck mass particle at the Planck scale  can be considered to be a
universe in itself [20]. We also use the fact alluded to that  a typical elementary particle
like the pion can be considered to be the result of $N \sim 10^{40}$ Planck
masses. Using this in (\ref{e1a}), we get $r \sim l$, the pion
Compton wavelength as required. Further, in this latter case, using (\ref{e2a}),i.e. $k_BT = kl^2/N$ and  (\ref{e3a}) and
(\ref{e4a}), we get for a pion, remembering that $m^2_P/N = m^2,$ 
$$k_ B T = \frac{m^3 c^4 l^2}{\hbar^2} = mc^2,$$
which of course is the well known formula for the Hagedorn temperature for
elementary particles like pions. In other words, this confirms the conclusions
in [1], that we can treat an elementary particle as a series of some
$10^{40}$ Planck mass oscillators. However it must be observed from 
(\ref{e2a}) and (\ref{e3a}), that while the Planck mass gives the highest
energy state, an elementary particle like the pion is in the lowest energy
state. This explains why we encounter elementary particles, rather than
Planck mass particles in nature. Infact as already noted [21,22], a Planck
mass particle decays via the Bekenstein radiation within a Planck time
$\sim 10^{-42}secs$. On the other hand, the lifetime of an elementary particle
would be very much higher.\\
\indent It must be mentioned that a fundamental minimum scale leads to violations of Lorentz symmetry - a topic that is being studied with much interest with a large body of literature [23,24,25]. The effects are in the nature of small corrections which hopefully will be detectable in ultra high energy cosmic rays, at extreme energies, though perhaps not in earth based colliders in the near future. In any case the efficacy of our above oscillator model can be seen by the fact that we recover correctly the masses and Compton scales in the order of magnitude sense and also get the correct Bekenstein and Hagedorn formulas as seen above, and get the correct estimate of the mass of the universe itself, as will be seen in the sequel. This is sufficient for the present purpose.\\
\indent It is interesting to note that the oscillator model has been used to give a formula for the mass spectrum of all known elementary particles [26]. It may be mentioned that Ng gives another interesting mass spectrum where important particles are associated with prime numbers [27]. Another important approach is that of Beck in which he uses stochastic quantization on chaotic strings to deduce low energy Bosonic and Fermionic masses (Cf.refs.[28,29]).\\
\indent Using the fact that the universe consists of $n \sim 10^{80}$ elementary
particles like the pions, the question is, can we think of the universe as
a collection of $10^{120}$ Planck mass oscillators? This is what we will now
show. Infact if we use equation (\ref{e1a}) with
$$N \sim 10^{120},$$
we can see that the extent $r \sim 10^{28}cms$ which is of the order of the diameter of the
universe itself. Next using (\ref{e4a}) we get
\begin{equation}
n \hbar \omega_0^{(min)} \frac{l_P}{10^{28}} \approx m_P c^2 \times 10^{60} \approx Mc^2\label{e5a}
\end{equation}
(where the multiplying factor on the left side viz., $l_P/r$ is dimensionless)
which gives the correct mass $M$, of the universe. We have started with (\ref{e4a}), taking $r$ to be the radius of the universe to get the lowest energy state of the oscillator system and then got in (\ref{e5a}) the whole energy for the $10^{120}$ oscillators. We get the same result if in (\ref{e4a}) we take $r$ to be the pion Compton wavelength and compute the energy for the $10^{80}$ pions in the universe. In other words the universe
itself can be considered to be a normal mode of Planck scale oscillators.\\
\indent It may be pointed out that the number $N \sim 10^{120}$ has also featured in the work of L. Nottale [30]. It is related to a scale dual to the minimum Planck scale, something which arises from the modification of the Uncertainty Principle due to the minimum scale [31,32,33,34,35,36].\\
\indent Finally it is interesting to note that equations like (\ref{e1a}), which give the famous Eddington formula and also leads to the well known so called large number empirical coincidences of Dirac have received some attention in recent times [37,38,39]. Interestingly all this has relevance to the recently discovered variation of the supposedly constant fine structure constant and related problems [40,4142].
\section{The Modified Uncertainty Principle}
As we noted above, It is well known that with the minimum space time interval, the Uncertainty Principle gets modified [43]. There is now an extra or second term:
\begin{equation}
\Delta x \approx \frac{\hbar}{\Delta p} + \alpha \frac{\Delta p}{\hbar}\label{e14}
\end{equation}
The first term gives the usual Heisenberg Uncertainty Principle. We will now relate the second term in (\ref{e14}) to large scale fluctuations $\sim \sqrt{N}$, in the universe, where $N \sim 10^{80}$ is the number of elementary particles encountered earlier. Infact the second term gives with $\Delta x = R$ the radius of the universe,
$$\frac{GmN}{c^2} = \frac{\Delta p}{h} l^2 = \frac{\sqrt{N}mc}{h}l^2$$
where the total uncertainty in momentum is $\sqrt{N}mc$. 
Whence we have
\begin{equation}
G = \frac{c^3l^2}{h\sqrt{N}}\label{e15}
\end{equation}
On the other hand without any reference to the Modified Uncertainty Principle, using the fluctuation in the particle number above we have [44]
\begin{equation}
G = \frac{lc^2}{m\sqrt{N}} = \frac{l^2c^3}{\hbar\sqrt{N}}\label{e16}
\end{equation}
It can immediately be seen that (\ref{e15}) is the same as (\ref{e16}).\\
In other words the second term, and therefore the spacetime fuzziness, is caused by the flouctuation in the number of particles in the universe. Indeed this was argued earlier from a different standpoint [45]. We could equivalently argue that the first term is related to electromagnetism (and Quantum Mechanics) and the second term to gravitation. For this we note that we have, for the gravitational energy
\begin{equation}
\frac{GNm^2}{R} = mc^2 (m = m_\pi\, \mbox{the \, pion \, mass}\,)\label{e17}
\end{equation}
while for the electromagnetic energy we have
\begin{equation}
\frac{e^2}{l_\pi} = mc^2 (m = m_e\, \mbox{the \, electron \, mass}\,)\label{e18}
\end{equation}
From (\ref{e17}) and (\ref{e18}) we have
\begin{equation}
\frac{Gmm^2}{e^2} = \frac{R}{Nl_\pi}\label{e19}
\end{equation}
If in (\ref{e19}) which is an equation in the ``Large Number'' spirit in which the distinction between the electron and pion masses gets blurred [46], we use the well known Eddington formula referred to earlier we get
\begin{equation}
\frac{G\sqrt{N}m^2}{e^2} \approx 1\label{e20}
\end{equation}
which is the emperically well known formula comparing the gravitational and electromagnetic field strengths.
\section{Remarks}
It is interesting to note in the context of the above Planck scale oscillator arrays, that in this case the energy is given by [2]
\begin{equation}
E = 2A cos kl\label{e21}
\end{equation}
where (\ref{e21}) is a base energy, $k$ the usual wave number (momentum) and $l$ is the spacing. What equation (\ref{e21}) shows is that we have
\begin{equation}
v = \frac{Ap}{\hbar^2}l^2\label{e22}
\end{equation}
as $l$ is small. From (\ref{e22}) we can conclude the following: If $A = mc^2$ then $l$ will be the Compton scale as $v$ approaches $c$. Alternatively $A$ or $v$, given the Compton (Planck) scale has the (minimum) scale, there is a maximal attainable velocity $c$, imposed by the lattice structure. Another interesting point is that in an equation like (\ref{e5a}), if we use instead of the Planck constant, a scaled Planck constant $H \sim 10^{93}$ as discussed extensively in the literature (Cf. for example [2]), then we recover consistently the frequency for oscillation of the universe,
\begin{equation}
\omega_u = \frac{mc^2}{H}\label{e23}
\end{equation}
Moreover from (\ref{e23}) the radius of the universe turns out to be the scaled Compton interval
$$R = \frac{H}{Mc}$$

\noindent {\large {\bf REFERENCES}}\\ \\
\noindent 1.  B.G Sidharth, Annales de la Fondation Louis de Broglie, 27 (2), 2002, pp.333ff.\\
\noindent 2. B.G. Sidharth, "Chaotic Universe: From the Planck to the Hubble Scale", Nova Science Publishers, Inc., New York, 2001.\\
\noindent 3. G. 't Hooft, ArXiv:gr-qc/9601014.\\
\noindent 4. T.D. Lee, "Particle Physics and Introduction to Field Theory", Harwood Academic, 1981, pp.383ff.\\
\noindent 5. B.G. Sidharth, ``Geometry and Quantum Mechanics'', to appear in Annales de la Fondation Louis de Broglie.\\
\noindent 6. M. Jacob (Ed.), ``Gauge Theory and Neutrino Physics'', North Holland, Amsterdam, 1978.\\
\noindent 7. B.G. Sidharth, Il Nuovo Cimento, 116B (6), 2001, pg.4 ff.\\
\noindent 8. B.G. Sidharth, Il  Nuovo Cimento, 117B (6), 703ff, 2002.\\
\noindent 9.  B.G. Sidharth, Found.Phys.Lett., August 2002.\\
\noindent 10. B.G. Sidharth, Found.Phys.Lett., 15 (6), 2002, pp.577-583.\\
\noindent 11. P.A.M. Dirac, in ``Monopoles in Quantum Field Theory'', Eds. N.S. Craigie et al., World Scientific, Singapore, 1982.\\
\noindent 12. B.G. Sidharth, Il Nuovo Cimento, 118B (1), 2003, pp.35-40.\\
\noindent 13. C.H. Ohanian, and R. Ruffini, "Gravitation and Spacetime", New York, 1994, p.397.\\
\noindent 14. P.G.Bergmann, ``Introduction to the Theory of
Relativity'', Prentice-Hall, New Delhi, 1969, 245ff.\\
\noindent 15. B.G. Sidharth, Gravitation and Cosmology, 4 (2) (14), 1998, p.158ff.\\
\noindent 16. B.G. Sidharth, Found.Phys.Lett., 16 (1), 2003, p.91-97.

\noindent 17.  B.G. Sidharth, {\it Found. of Phys.Lett.} {\bf 15} (6), December (2002), 577-583.\\
\noindent 18. Y. Jach NG and H. Van Dam, {\it Mod.Phys.Lett.A.} {\bf 9} (4), (1994), p.335-340.\\
\noindent 19.  D.L. Goodstein, {\it States of Matter} (Dover Publications Inc., New York, 1985), p.160ff.\\
\noindent 20.  N. Rosen, {\it Int.J.Th.Phys.} {\bf 32} (8), (1993), p.1435-1440.\\
\noindent 21. B.G. Sidharth, {\it Chaotic Universe: From the Planck to the Hubble Scale} (Nova Science Publishers Inc., New York, 2001), p.65ff.\\
\noindent 22. B.G. Sidharth, {\it Chaos Solitons and Fractals} {\bf 12} (1), (2000), 173-178.\\ 
\noindent 23. A.V.  Olinto, {\it Phys.Rep.} {\bf 333-334}, (2000), pp.329-348.\\
\noindent 24. S. Coleman and S.L. Glashow, {\it Phys.Rev.D.} \underline{59}, (1999), 116008.\\
\noindent 25. B.G. Sidharth, ``Non Commutative Geometry and Issues'' Physics/0312109.\\
\noindent 26. B.G. Sidharth, ``A Formula for the Mass Spectrum of Elementary Particles'', to appear in Hadronic Journal.\\
\noindent 27. Sze Kui Ng, ``A Computation of the Mass spectrum of Mesons and Baryons'', hep-ph/0208098.\\
\noindent 28. C. Beck, {\it Spatio-temporal Chaos and vacuum Fluctuations of Quantized Fields}, Advances in Nonlinear Dynamics, {\bf vol.21} (World Scientific, Singapore, 2002); hep-th/0207081.\\
\noindent 29. C. Castro, {\it Chaos, Solitons and Fractals} {\bf 15},(2003), 797.\\
\noindent 30. L. Nottale, {\it Fractal Spacetime and Microphysics, Towards Scale Relativity} (World Scientific, Singapore, 1992).\\
\noindent 31. B.G. Sidharth, {\it Chaos, Solitons and Fractals} {\bf 15},(2003), 593-595.\\
\noindent 32. D. Gross and P. Mende, {\it Nucl.Phys.} {\bf B303}, (1988), 407.\\
\noindent 33. D. Amati, M. Ciafaloni and G. Veneziano, {\it Phys.Lett.} {\bf B216}, (1989), 41.\\
\noindent 34. C. Castro, {\it Foundations of Physics,} {\bf 30},(2000), 1301.\\
\noindent 35. C. Castro, {\it Chaos, Solitons and Fractals,} {\bf 11}, (2000), 1721.\\
\noindent 36. C. Castro, {\it Chaos, Solitons and Fractals,} {\bf 12}, (2001), 1585.\\
\noindent 37. B.G. Sidharth, {\it Int.J.Mod.Phys.A.} {\bf 13} (15), (1998), p.2599ff.\\
\noindent 38. M. Kafatos, S. Roy, R. Amoroso., {\it Scaling in Cosmology and the Arrow of Time} in Studies in the Structure of Time. Edited by R. Buccheri and M. Saniga (Kluwer Academic Publishers, New York, 2000), pages 191-200.\\
\noindent 39. J. Glanz: {\it Science} {\bf 22}, (1998), 2156.\\
\noindent 40. B.G. Sidharth, {\it Il Nuovo Cimento,} {\bf 115B} (12), (2), (2000), pg.151.\\
\noindent 41. J.P. Uzan,  ``The Fundamental constants and their variations:Observational status and theoretical motivations'' he-ph/0012539.\\
\noindent 42. Y. Jack Ng., ``Selected topics in Planck scale Physics'' hep-th/0305019.\\ 
\noindent 43. B.G. Sidharth, Chaos, Solitons and Fractals, 15, 2003, p.593-595.\\
\noindent 44. B.G. Sidharth, Int.J. of Mod.Phys.A 13(15), 1998, pp2599ff.\\
\noindent 45. B.G. Sidharth, Chaos, Solitons and Fractals, 16 (4), 2003, p.613-620.\\
\noindent 46.  C.W. Misner, K.S. Thorne and J.A. Wheeler, "Gravitation",
W.H. Freeman, San Francisco, 1973, pp.819ff.
\end{document}